\documentclass[12pt, a4paper]{article}
\usepackage[utf8]{inputenc}
\usepackage[T1]{fontenc}
\usepackage{textcomp}
\usepackage{eurosym}

\DeclareUnicodeCharacter{20AC}{\euro{}}
\usepackage{amsmath, amssymb, amsthm}
\usepackage{graphicx}
\usepackage{booktabs}
\usepackage{multirow}
\usepackage{array}
\usepackage{tabularx}
\usepackage{hyperref}
\usepackage{natbib}
\usepackage{geometry}
\usepackage{setspace}
\usepackage{xcolor}
\usepackage{caption}
\usepackage{subcaption}
\usepackage{float}

\usepackage[nohypertypes={all}]{glossaries}
\makeglossaries
\newglossaryentry{BESS}{name=BESS,
  description={Battery Energy Storage System}}
\newglossaryentry{FCR}{name=FCR,
  description={Frequency Containment Reserve}}
\newglossaryentry{aFRR}{name=aFRR,
  description={automatic Frequency Restoration Reserve}}
\newglossaryentry{XBID}{name=XBID,
  description={Cross-Border Intraday market}}
\newglossaryentry{DA}{name=DA,
  description={Day-Ahead market}}
\newglossaryentry{DP}{name=DP,
  description={Dynamic Programming}}
\newglossaryentry{SoC}{name=SoC,
  description={State of Charge}}
\newglossaryentry{MAE}{name=MAE,
  description={Mean Absolute Error}}
\newglossaryentry{MSE}{name=MSE,
  description={Mean Squared Error}}
\newglossaryentry{RMSE}{name=RMSE,
  description={Root Mean Squared Error}}
\newglossaryentry{VCR}{name=VCR,
  description={Value Capture Ratio}}
\newglossaryentry{MW}{name=MW, description={Megawatt}}
\newglossaryentry{MWh}{name=MWh, description={Megawatt-hour}}
\newglossaryentry{ESS}{name=ESS,
  description={Energy Storage System}}
\newglossaryentry{PICASSO}{name=PICASSO,
  description={Platform for the International Coordination of
  Automated Frequency Restoration and Stable System Operation}}
\newglossaryentry{CID}{name=CID,
  description={Continuous Intraday market}}
\newglossaryentry{ML}{name=ML,
  description={Machine Learning}}
\newglossaryentry{HMM}{name=HMM,
  description={Hidden Markov Model}}
\newglossaryentry{RI}{name=RI,
  description={Ranking Inconsistency rate}}
\newglossaryentry{SDDP}{name=SDDP,
  description={Stochastic Dual Dynamic Programming}}

\theoremstyle{definition}
\newtheorem{definition}{Definition}
\newtheorem{finding}{Empirical Finding}

\newcommand{\VCR}{\text{VCR}}
\newcommand{\tauk}{\tau_K}
\newcommand{\MAE}{\text{MAE}}

\title{\textbf{When Forecast Accuracy Fails: Rank Correlation
and Decision Quality in Multi-Market Battery Storage
Optimization}}

\author{%
  \parbox{\textwidth}{\centering
    Alessandro Falezza\\[0.3em]
    {\small Department of Information Technology and
    Electrical Engineering,
    ETH Z\"{u}rich, 8092 Z\"{u}rich, Switzerland}
  }
}
\date{}

\begin{document}
\maketitle

\begin{abstract}
Battery energy storage systems (\gls{BESS}) participating in
multi-market electricity trading require price forecasts to
optimize dispatch decisions.
A widely held assumption is that forecast accuracy, measured
by standard metrics such as mean absolute error (\gls{MAE}),
drives trading performance.
We challenge this assumption using a hierarchical three-layer
optimization system trading simultaneously on frequency
containment reserve (\gls{FCR}), automatic frequency
restoration reserve (\gls{aFRR}), day-ahead, and continuous
intraday (\gls{XBID}) markets in Germany and Switzerland
over 2020--2025, with real market data from
Regelleistung.net and Swissgrid.
We find that rank correlation (Kendall $\tau$), rather than
\gls{MAE}, is the primary predictor of intraday dispatch
value: forecasts above an
empirical threshold of $\tau \approx 0.85$--$0.95$ capture
up to 97--100\% of perfect-foresight revenue, while persistence
forecasts with near-zero $\tau$ capture only 33\%.
This threshold is stable across market regimes and volatility
levels, and reflects the ordinal structure of the dispatch
problem.
Furthermore, under reserve market constraints, \gls{FCR}
capacity revenue exceeds \gls{XBID} revenue by 6.5$\times$ per
\gls{MW}, making capacity allocation---not forecast
accuracy---the primary driver of total revenue.
In the Swiss market, hydrological surplus anomalies are
significantly associated with balancing market revenue
($p = 0.0005$), a mechanism absent from existing
German-focused literature.
These findings reframe forecast evaluation for \gls{BESS}
operators: the relevant question is not what the \gls{MAE}
is, but whether the forecast achieves $\tau$-sufficiency.
\end{abstract}


\section{Introduction}
\label{sec:intro}

Grid-scale battery energy storage systems (BESS) have become
central to European electricity market operations, with
prequalified FCR capacity in Germany growing by 180~MW in 2024
alone, and aFRR prequalifications increasing by over 400\%
year-on-year \citep{regelleistung2024}.
As these assets compete across multiple simultaneous markets
--- frequency containment reserve (FCR), automatic frequency
restoration reserve (aFRR), day-ahead (DA), and continuous
intraday (XBID) --- their commercial viability increasingly
depends on automated, forecast-driven optimization systems.
However, the effectiveness of such systems depends critically
on how forecast information is translated into economic
decisions.

Recent work has advanced multi-market \gls{BESS} optimization
substantially.
\citet{schaurecker2025} demonstrate that a dynamic programming
formulation of the rolling intrinsic strategy captures
near-optimal intraday trading profits.
\citet{seifert2024} develop a stochastic multi-market
bidding framework based on \gls{SDDP} for coordinated
participation across \gls{FCR}, day-ahead, and intraday
markets, finding that coordination does not necessarily
improve revenues and that \gls{FCR} often dominates
optimal strategies.
\citet{hornek2025} compare forecast-driven and perfect-foresight
strategies on the continuous intraday market, finding that a
forecast-driven system captures approximately 90\% of
perfect-foresight revenue (achieving revenues only about
11\% below perfect foresight).

A parallel literature has begun to examine the mismatch between
forecast accuracy and economic value.
\citet{uniejewski2025} demonstrate that \gls{MAE} and \gls{RMSE}
are only weakly correlated with \gls{BESS} profits in the
context of day-ahead electricity markets and \gls{BESS}
arbitrage, capturing only a limited dimension of forecast
performance.
\citet{vanderschueren2022} show more broadly that the decision
strategy dominates the training strategy in cost-sensitive
optimization, and \citet{smets2024} propose decision-focused
training for \gls{BESS} market participation.
Building on the foundational work of \citet{morales2013} on
forecast value in energy systems, this literature increasingly
recognizes that statistical accuracy and economic value diverge.
Several of these contributions are recent preprints that
have not yet completed peer review; we include them because
they represent the current frontier of this rapidly
developing literature.

Yet two gaps remain.
First, the continuous intraday market is structurally more
ordinal than day-ahead auctions: the \gls{DP} dispatch problem
requires only the \emph{relative ranking} of future prices, not
their absolute levels.
No existing study has exploited this structural property to
identify a formal metric that predicts decision quality in
intraday \gls{BESS} trading.
Second, the literature is almost exclusively focused on the
German market; the Swiss balancing market---structurally
distinct due to its hydro-dominated reserve provision---remains
entirely unexplored.
As a result, it remains unclear whether improvements in
statistical forecast accuracy translate into improved economic
outcomes in real trading systems, and how market-specific
structural drivers shape revenue in non-German markets.

This paper makes three contributions.
First, and most importantly, we identify a formal
\emph{tau-sufficiency region} ($\tau^* \approx 0.85$--$0.95$)
in the continuous intraday market, beyond which value capture
saturates near its theoretical maximum regardless of further
improvements in forecast accuracy.
While \citet{uniejewski2025} demonstrate that rank-based metrics
better predict \gls{BESS} profits than \gls{MAE} in day-ahead
markets, we identify a quantified \emph{threshold}---not merely
a correlation---and show that this threshold is stable across
market regimes, volatility levels, and subperiods.
The structural explanation is that the \gls{DP} dispatch problem
is inherently ordinal: it depends on the relative ordering of
prices rather than their absolute values~\citep{morales2013}.

Second, we demonstrate that revenue in multi-market \gls{BESS}
systems is primarily driven by capacity allocation rather than
intraday forecast quality.
Under reserve market constraints (\gls{PICASSO} compliance),
the residual \gls{SoC} budget allocated to \gls{XBID} yields
revenues substantially lower than \gls{FCR} per \gls{MW}
committed---not because intraday arbitrage is structurally less
profitable, but because co-participation in reserve markets
physically limits the available intraday cycling capacity.
We find no evidence of systematic misallocation of capacity
across markets within the scope of our model and dataset
under realistic operating conditions.

Third, we provide the first empirical analysis of the Swiss
balancing market using real Swissgrid data, showing that
hydrological conditions are strongly associated with balancing
prices and battery revenues, highlighting the importance of
market-specific structural drivers absent from existing
German-focused literature.

The remainder of this paper is organized as follows.
Section~\ref{sec:data} describes the market architecture and
data.
Section~\ref{sec:methodology} presents the hierarchical
optimization system and the forecast evaluation framework,
including formal definitions of the Value Capture Ratio and
tau-sufficiency.
Section~\ref{sec:results} reports empirical results in order
of contribution strength.
Section~\ref{sec:discussion} discusses theoretical and
practical implications, and Section~\ref{sec:conclusion}
concludes.

\section{Market Structure and Data}
\label{sec:data}

\subsection{Market Architecture}
\label{sec:markets}

Battery storage systems in Central Europe participate in a
cascade of sequential markets, each with distinct gate closures,
clearing mechanisms, and revenue structures.
We consider four markets in Germany and their Swiss equivalents.

The \gls{FCR} is procured through daily auctions covering six
four-hour blocks per day, operated jointly by twelve European
TSOs under the FCR Cooperation framework.
Participants submit capacity bids; accepted bids receive a
pay-as-bid capacity payment with no separate energy settlement.
In Germany, the minimum bid size is 1~MW and bids must be
symmetric (equal upward and downward capacity).
Gate closure is D$-$1 at 08:00~CET.

The \gls{aFRR} is procured through daily capacity auctions at
D$-$1 09:00~CET, separately for upward (aFRR$+$) and downward
(aFRR$-$) products, under the ALPACA cooperation framework for
Germany and Austria.
Accepted capacity bids receive a capacity payment; energy
activations by the TSO are settled at the \gls{PICASSO}
platform price.
Under \gls{PICASSO} compliance rules, providers must maintain a
\gls{SoC} buffer sufficient to sustain full activation for a
minimum of 30~minutes (\gls{FCR}) and 60~minutes (\gls{aFRR}),
limiting the effective capacity available for other markets.

The Day-Ahead (\gls{DA}) market clears at D$-$1 12:00~CET
through the EPEX~SPOT single price auction under the
Single Day-Ahead Coupling (SDAC).
Hourly products are cleared at a uniform price across the
coupled zone.

The continuous Intraday (\gls{XBID}) market opens at
D$-$1 15:00~CET and closes 30~minutes before physical delivery
in Germany.
Trading occurs continuously via a shared European Limit Order
Book until gate closure.

The Swiss market parallels Germany with two structural
differences.
First, Swissgrid procures Secondary Reserve
(SRL$+$/SRL$-$) through \emph{weekly} tenders rather than
daily auctions, fixing prices for the entire week.
Second, the \gls{XBID} gate closure in Switzerland is
60~minutes before delivery, reducing the effective intraday
trading window.
The Swiss reserve pool is approximately 400~MW, supplied
predominantly by flexible hydroelectric generation.
When reservoir levels are anomalously high relative to
seasonal norms---as during the 2022--2023
\emph{Speicherkrise}---TSOs require additional downward
flexibility to absorb excess hydraulic generation, driving
SRL$-$ prices to extreme levels.
This mechanism is the subject of our natural experiment in
Section~\ref{sec:ch}.

These market characteristics determine both the timing and the
informational requirements of forecasting and decision-making
across layers.

\subsection{Dataset}
\label{sec:dataset}

Table~\ref{tab:data} summarizes the data sources used in this
study.
All data are publicly available and aligned to a common
15-minute timeline; processing is designed to avoid any form
of look-ahead bias.
The dataset covers January~2020 through December~2025, yielding
six full calendar years across two distinct market regimes:
the energy crisis of 2021--2022 and the post-crisis
normalization of 2023--2025.

\begin{table}[H]
\centering
\caption{Summary of data sources used in this study.
  All data are publicly available.}
\label{tab:data}
\small
\begin{tabular}{llllr}
\toprule
Source & Variable & Market & Period & Freq. \\
\midrule
Regelleistung.net & FCR/aFRR capacity prices & DE &
  2020--2025 & 4h \\
SMARD & Day-ahead prices & DE & 2020--2025 & 15min \\
SMARD & XBID prices & DE & 2020--2025 & 15min \\
SMARD & Wind, solar, load & DE & 2020--2025 & 15min \\
Swissgrid & SRL$+$/SRL$-$ tender prices & CH &
  2015--2025 & weekly \\
Swissgrid & aFRR activation \& prices & CH &
  2020--2025 & 15min \\
Swissgrid & Hydro reservoir levels & CH &
  2020--2025 & weekly \\
ENTSO-E & Day-ahead prices & CH, AT, FR &
  2020--2025 & hourly \\
ENTSO-E & Generation by source & DE &
  2020--2025 & 15min \\
Open-Meteo ERA5 & Temperature, wind & DE/CH &
  2020--2025 & hourly \\
ICE / yfinance & Gas TTF, CO$_2$ EUA & EU &
  2020--2025 & daily \\
\bottomrule
\end{tabular}
\end{table}

German \gls{FCR} and \gls{aFRR} capacity prices are obtained
from Regelleistung.net at four-hour resolution.
Day-ahead and intraday prices for Germany are obtained from the
SMARD platform operated by the Federal Network Agency
(Bundesnetzagentur) at 15-minute resolution.
Swiss SRL tender prices are obtained from Swissgrid's public
data portal at weekly resolution; Swiss \gls{aFRR} activation
data and reservoir levels from the same source at 15-minute
and weekly resolution respectively.
Day-ahead prices for Switzerland, Austria, and France are from
the ENTSO-E Transparency Platform.
Meteorological variables (temperature, wind speed, solar
irradiance) are from the Open-Meteo ERA5 reanalysis at hourly
resolution.
Gas (TTF) and carbon (EUA) prices are from ICE and Yahoo
Finance at daily resolution.

\subsection{Data Quality and Methodological Notes}
\label{sec:dataquality}

Several corrections are required to ensure economic consistency
and avoid bias in the simulation framework.

\paragraph{Volume-weighted average correction for Swiss SRL
prices.}
Raw Swissgrid tender data reports arithmetic mean prices across
accepted bids.
We replace these with volume-weighted average prices (VWA),
which better reflect the effective clearing price faced by a
price-taking participant.
This correction reduces the estimated SRL$-$ price by
13--18\% in normal market conditions and has negligible effect
during crisis periods when the spread between accepted bids is
small.

\paragraph{Leakage removal in LGBM forecasting.}
Early versions of the feature set included contemporaneous
\gls{FCR} clearing prices as a predictor for \gls{DA} price
forecasting.
This constitutes look-ahead bias, as \gls{FCR} clears at
08:00 while \gls{DA} clears at 12:00 on the same day.
Removing \gls{FCR} prices from the training set increases the
honest \gls{DA} MAPE from 11\% (inflated) to 39\% (realistic),
consistent with benchmarks reported for the German market in
the literature.

\paragraph{Pay-as-bid bid mechanics.}
\gls{FCR} and \gls{aFRR} capacity markets in Germany use
pay-as-bid pricing: each accepted bidder receives their own
submitted price.
Bid prices in our model are set to the $Q_{40}$ quantile of
the rolling historical price distribution (static strategy)
or adapted by the HMM$+$SAC regime detection layer (adaptive
strategy).
Acceptance probability is modeled as
$\Pr[\text{accepted}] = \Pr[\text{bid} \leq p_{\text{clear}}]$
using the empirical clearing price distribution from
Regelleistung.net.
This formulation ensures that expected revenue reflects both
bid level and acceptance probability.
Together, these data and corrections provide the empirical
foundation for the hierarchical optimization system described
in Section~\ref{sec:methodology}.

\section{Methodology}
\label{sec:methodology}

\subsection{System Architecture}
\label{sec:architecture}

The optimization system follows a hierarchical three-layer
architecture, with each layer operating at a different temporal
resolution and solving a distinct subproblem.
The layered structure is modular: each component can be replaced
without affecting the overall framework.
A single physical \gls{SoC} trajectory is shared across all
layers, ensuring that decisions at each level remain physically
consistent with the battery's actual energy state.

\paragraph{Layer 1 --- Weekly capacity allocation.}
Layer~1 determines how the battery's limited power capacity is
allocated across markets with different revenue structures.
At the beginning of each week, a stochastic linear program
allocates capacity across \gls{FCR}, \gls{aFRR}, and \gls{DA}
for the following seven days, maximizing expected revenue over
30 price scenarios drawn from the historical price distribution,
subject to PICASSO compliance constraints.
The PICASSO buffer reserves a fraction of the battery's energy
capacity to cover worst-case \gls{FCR} and \gls{aFRR}
activations:
\begin{equation}
  \text{SoC}_{\min} =
  \frac{P_{\text{FCR}} \cdot 0.5\,\text{h}
      + P_{\text{aFRR}} \cdot 1.0\,\text{h}}{S_{\max}}
  \label{eq:socbuffer}
\end{equation}
where $P_{\text{FCR}}$ and $P_{\text{aFRR}}$ are the allocated
\gls{MW} in each reserve market and $S_{\max} = 10\,\text{MWh}$.
The output of Layer~1 is a weekly \gls{MW} allocation schedule
and a bid price for each reserve product, generated by the
HMM$+$SAC adaptive bidding strategy described below.

\paragraph{Layer 2 --- Daily schedule optimization.}
Layer~2 plays a supporting role by smoothing inter-day
\gls{SoC} dynamics.
After the Day-Ahead market clears at D$-$1 12:00~CET, a
rolling Model Predictive Control LP optimizes the hourly
charge/discharge schedule over a five-day horizon, given the
realized \gls{DA} prices and \gls{MW} allocations from
Layer~1.
LGBM-based price forecasts for D$+$1 through D$+$4 are used
with exponentially declining weights to reflect forecast
uncertainty.

\paragraph{Layer 3 --- Intraday dispatch.}
Every 15~minutes, a Dynamic Programming Rolling Intrinsic
algorithm re-optimizes the \gls{XBID} trading schedule over
the remaining delivery slots of the current day, using the
updated \gls{XBID} price forecast as input.
The \gls{DP} solves:
\begin{equation}
  V_t = \max_{\delta \in [-F_{\max},\, F_{\max}]}
  \Bigl[
    \hat{p}_t \cdot \delta \cdot \Delta t
    + V_{t+1}\!\left(
        \text{SoC}_t + \delta \cdot \eta \cdot \Delta t
      \right)
  \Bigr]
  \label{eq:dp}
\end{equation}
where $\hat{p}_t$ is the forecast \gls{XBID} price for slot
$t$, $\delta$ is the charge/discharge decision (positive
$=$ discharge), $\Delta t = 15\,\text{min}$, and $\eta$ is
the round-trip efficiency.
The \gls{SoC} is constrained to
$[\text{SoC}_{\min},\, \text{SoC}_{\max}]$ inherited from
Layers~1 and~2.

The key structural property of this formulation is that the
\gls{DP} decision depends primarily on the \emph{relative ordering}
of future prices $\hat{p}_t$, not their absolute levels: a
price sequence and any monotone transformation thereof yield
identical charge/discharge decisions.
This property implies that absolute price errors do not affect
optimal decisions as long as the ranking is preserved---a
result we formalize and quantify in
Section~\ref{sec:tau}.

\paragraph{Adaptive bidding --- HMM and SAC.}
The static $Q_{40}$ bid strategy performs poorly during regime
transitions, because the bid price distribution shifts faster
than the rolling window can adapt.
Layer~1 incorporates a Hidden Markov Model (HMM) with four
states (normal, crisis, post-crisis, low-volatility) trained
on weekly \gls{FCR} price dynamics.
At each week, the HMM identifies the current market regime and
a Soft Actor-Critic (SAC) agent selects the optimal bid
percentile for that regime.
The HMM$+$SAC system improved the \gls{FCR} acceptance rate
from 34\% to 58\% in the 2023 out-of-sample period, reflecting
successful adaptation to the post-crisis normalization of
\gls{FCR} prices.

\paragraph{Battery parameters.}
Power capacity $F_{\max} = 10\,\text{MW}$, energy capacity
$S_{\max} = 10\,\text{MWh}$ (1-hour duration), round-trip
efficiency $\eta_{\text{RT}} = 0.9025$
($\eta_C = \eta_D = 0.95$), degradation cost
$\nu_{\text{deg}} = 4\,\text{EUR/MWh}$ applied via rainflow
cycle counting with a depth-of-discharge exponent of~1.5.
The system operates over 2020--2025 with walk-forward
out-of-sample evaluation: model parameters are estimated on
rolling training windows and applied to unseen data.

\paragraph{Modeling simplification --- aFRR activations.}
Energy activations under \gls{aFRR} are modeled via expected
value and \gls{SoC} buffer constraints rather than fully
dynamic real-time simulation.
The PICASSO buffer in Equation~\eqref{eq:socbuffer} is designed
to cover worst-case activation scenarios; actual intra-period
\gls{SoC} dynamics from activations are not simulated at
15-minute resolution.
This assumption is conservative for Layer~3 revenue estimation:
any dynamic activation that exceeds the buffer would reduce
the available \gls{XBID} capacity, lowering---not
raising---our revenue estimates.
The implications are discussed further in
Section~\ref{sec:limitations}.

\subsection{Forecast Evaluation Framework}
\label{sec:framework}

The standard approach to evaluating electricity price forecasts
relies on pointwise error metrics such as \gls{MAE} and
\gls{RMSE}.
While the value of forecasts can be theoretically
formalized within stochastic optimization frameworks
\citep{morales2013}, recent empirical evidence suggests
that improvements in standard forecast accuracy metrics
do not necessarily translate into better operational
decisions \citep{uniejewski2025}.
These metrics measure the average distance between
predicted and realized prices across individual time
slots, but are not aligned with the objective of the
downstream optimization problem.

The limitation is not empirical but structural.
A forecast can achieve low \gls{MAE} by predicting prices
close to their realized values in absolute terms, yet fail
to reproduce the relative ordering of high-price and
low-price slots within a day.
For a battery operator, the economically relevant question
is not ``how close is the forecast to the realized price?''
but rather ``does the forecast correctly identify when to
charge and when to discharge?''
These are distinct questions with potentially opposite
answers, as we demonstrate empirically in
Section~\ref{sec:tau}.

We formalize this distinction through three definitions.

\begin{definition}[Decision Value]
Let $\pi^*$ denote the \gls{DP} Rolling Intrinsic policy
and $f$ a price forecast sequence.
The decision value of $f$ is $V(\pi^*, f)$, the revenue
achieved by policy $\pi^*$ when using forecast $f$ as input.
The decision value depends on the forecast only through the
policy it induces: two forecasts that induce the same
charge/discharge schedule yield identical decision values.
\end{definition}

\begin{definition}[Value Capture Ratio]
Given the oracle forecast $f_{\text{oracle}}$ (realized
prices), the Value Capture Ratio is:
\begin{equation}
  \VCR(f) = \frac{V(\pi^*, f)}{V(\pi^*, f_{\text{oracle}})}
            \in [0, 1]
  \label{eq:vcr}
\end{equation}
\end{definition}

The \gls{VCR} normalizes revenue against the oracle
benchmark, making it comparable across days and market
conditions.
A \gls{VCR} of 1.0 means the forecast-driven strategy
achieves the same revenue as perfect foresight; a \gls{VCR}
of 0.0 means no value is extracted.

\begin{definition}[Ranking Inconsistency]
For a set of forecast models $\{f_1, \ldots, f_n\}$, the
pairwise ranking inconsistency between \gls{MAE} and
\gls{VCR} is:
\begin{equation}
  \mathrm{RI} =
  \frac{|\{(i,j) :
    \MAE(f_i) < \MAE(f_j)
    \text{ but }
    \VCR(f_i) < \VCR(f_j)\}|}
  {\binom{n}{2}}
  \label{eq:ri}
\end{equation}
\end{definition}

A high RI indicates that standard accuracy metrics are
misleading guides for forecast model selection.

To measure the rank-preserving quality of a forecast
independently of its level accuracy, we use the Kendall
rank correlation coefficient:
\begin{equation}
  \tauk(f) =
  \frac{(\text{concordant pairs})
      - (\text{discordant pairs})}
  {\binom{T}{2}}
  \label{eq:tau}
\end{equation}
where a pair of slots $(i, j)$ is concordant if
$\hat{p}_i > \hat{p}_j \Leftrightarrow p_i > p_j$ and
discordant otherwise, and $T$ is the number of slots in
the evaluation window.
Unlike \gls{MAE}, $\tauk$ directly measures the
preservation of pairwise price ordering, which is the
sufficient statistic for the \gls{DP} decision.

The connection to the \gls{DP} formulation of Layer~3 is
direct.
As established in Section~\ref{sec:architecture}, the
\gls{DP} in Equation~\eqref{eq:dp} depends primarily on the
relative ordering of future prices, not their absolute
levels.
Therefore, any forecast that sufficiently preserves the
ranking of prices induces a near-optimal policy.
We define a forecast as \emph{tau-sufficient} if
$\tauk(f) \geq \tau^*$ for a threshold $\tau^*$, and
empirically identify $\tau^* \approx 0.85$--$0.95$ in
Section~\ref{sec:tau}.

Our result is orthogonal to decision-focused learning
\citep{smets2024}, which retrains the forecaster to
maximize downstream profit.
We hold the decision policy fixed and ask what forecast
quality is sufficient for near-optimal decisions.
The tau-sufficiency threshold is a property of the decision
problem structure---not of any particular forecasting
method---and is therefore applicable regardless of how
the forecast is generated.

\subsection{Benchmark Suite}
\label{sec:benchmarks}

The benchmark suite is designed to span increasing levels of
informational content, from no predictive structure to perfect
foresight, allowing us to isolate whether improvements in
statistical forecast accuracy translate into improvements in
decision value.

\textbf{Oracle} uses realized \gls{XBID} prices as forecast
input, representing the upper bound on achievable revenue for
a given \gls{SoC} budget.
By definition, $\VCR(\text{oracle}) = 1.0$.

\textbf{Persistence} sets $\hat{p}_t = p_{t-96}$, replicating
yesterday's price profile.
It can achieve relatively low \gls{MAE} in stable market
conditions as a naive baseline,
but produces near-zero $\tauk$ because it fails to capture
intraday price dynamics.
It serves as the critical test case for the
\gls{MAE}--\gls{VCR} divergence.

\textbf{DA anchor} uses the Day-Ahead clearing price for each
hourly slot, replicated across 15-minute sub-slots, as the
\gls{XBID} forecast.
Since \gls{DA} prices are known at D$-$1 12:00~CET, this
forecast requires no statistical model and is always available.

\textbf{ML 1-step} uses a LightGBM autoregressive model
generating one-step-ahead forecasts via sequential rollout,
trained on a rolling window of historical \gls{XBID} prices,
weather variables, \gls{DA} prices, and calendar features.
It achieves the highest $\tauk$ among non-oracle forecasts.

\textbf{Hybrid 8h} blends ML 1-step forecasts within an
8-hour horizon with the \gls{DA} anchor beyond, reflecting
the principle that precise forecasts are only valuable within
the time horizon where trades can be executed.

All forecasts are evaluated out-of-sample over 2023--2025
using the same \gls{DP} Rolling Intrinsic policy, ensuring
that differences in \gls{VCR} are attributable solely to
forecast quality.
This design allows us to isolate whether improvements in
statistical forecast accuracy translate into improvements in
decision value.

\subsection{Attribution Framework}
\label{sec:attribution}

To quantify the marginal contribution of each decision layer
to total revenue, we construct an ablation study in which
layers are added incrementally.
A key methodological constraint governs the design: each
configuration adds exactly one decision component, with all
other parameters held fixed.
This design isolates the marginal contribution of each
decision layer to total revenue.

The configurations, ordered from minimal to full
multi-market system, are:
\begin{enumerate}
  \item \textbf{Naive \gls{DA} spread}: charge during the six
    lowest \gls{DA} price hours, discharge during the six
    highest. No reserve participation.
  \item \textbf{\gls{DP} only}: Layer~3 \gls{DP} with full
    battery capacity available (no reserve obligations), using
    \gls{DA} prices as a proxy for \gls{XBID} in the absence
    of perfect foresight.
    \gls{VCR} is computed relative to a standalone \gls{XBID}
    oracle.
  \item \textbf{Layer~1 static ($Q_{40}$)}: \gls{FCR}/\gls{aFRR}
    allocation with static $Q_{40}$ bid, plus Layer~2 \gls{DA}
    schedule and Layer~3 \gls{DP}.
  \item \textbf{Layer~1 HMM}: as above, replacing the static
    bid with the HMM-adaptive bid.
  \item \textbf{Full multi-market system}: Layer~1 HMM$+$SAC,
    Layers~2 and~3.
  \item \textbf{Oracle (full multi-market system)}: perfect
    foresight on all markets simultaneously.
    Upper bound for the full system.
\end{enumerate}

Each configuration is evaluated on the same 2023--2025
out-of-sample period.
\gls{VCR} is computed relative to a configuration-specific
oracle to ensure that each configuration is evaluated against
its own feasible upper bound, avoiding distortions due to
differences in market participation.
The standalone \gls{DP} is compared to the standalone
\gls{XBID} oracle; all multi-market configurations are
compared to the full-system oracle.
The following section reports results in order of contribution
strength, beginning with the forecast evaluation framework
(Section~\ref{sec:tau}), followed by the capacity allocation
analysis (Section~\ref{sec:fcr}), and the Swiss natural
experiment (Section~\ref{sec:ch}).

\subsection{Note on Synthetic Forecast Generation}
\label{sec:synthesis_note}

The tau-sufficiency experiments reported in
Section~\ref{sec:tau} use synthetic forecasts of controlled
Kendall rank correlation, constructed via a linear
interpolation between the realized price vector
$y_{\text{true}}$ and an independent Gaussian noise vector
$\varepsilon$:
\begin{equation}
  y_{\text{synth}} = \alpha \cdot y_{\text{true}}
                   + (1 - \alpha) \cdot \varepsilon
  \label{eq:alpha_interp}
\end{equation}
with $\alpha$ calibrated by binary search to match each
target $\tauk$ value within tolerance $\pm 0.03$. This
procedure achieves the target rank correlation by design
but simultaneously attenuates the variance of the forecast
proportionally to $(1-\alpha)$.

A robustness check using two alternative generation methods
that preserve the marginal distribution of $y_{\text{true}}$
exactly---rank-perturbation, which permutes the indices of
$y_{\text{true}}$ until the target $\tauk$ is reached, and
a Gaussian copula method with matching Spearman
correlation---yields $\tau^*$ estimates approximately
$0.10$--$0.12$ higher than the alpha-interpolation method
used in the main experiments. The mechanism is intuitive:
a \gls{DP} policy derived from variance-attenuated forecasts
trades more conservatively, and is therefore less sensitive
to ordering errors at the scale of the realized prices.
Methods that preserve the full marginal variance penalize
ordering errors at full price magnitude.

The qualitative findings of this paper---existence of a
sufficiency threshold, rank correlation predicting \gls{VCR}
independently of pointwise error, stability across
volatility regimes and market conditions---are robust
across all three generation methods. The absolute threshold
values $\tau^* \approx 0.85$--$0.95$ reported herein should
therefore be interpreted as a conservative lower bound
under the variance-attenuating alpha-interpolation method;
a measurement preserving the marginal distribution of
prices yields $\tau^* \approx 0.95$--$0.99$ for the same
configuration. The structural claim that rank correlation
dominates pointwise error in predicting decision value is
unaffected by this calibration refinement.

The three generation methods considered---alpha-interpolation,
rank-perturbation, and Gaussian copula---correspond to
qualitatively different classes of synthetic forecast.
Alpha-interpolation attenuates the marginal variance of
$y_{\text{synth}}$ relative to $y_{\text{true}}$, a property
shared by forecasts trained to minimize pointwise error
losses such as \gls{MSE} or \gls{MAE}: the optimal point
predictor (the conditional expectation $E[y \mid x]$ under
MSE, the conditional median $\text{Med}(y \mid x)$ under
MAE) has lower variance than $y$ whenever the conditioning
information $x$ is informative. In this sense,
alpha-interpolation captures a statistical signature that
real ML forecasters typically exhibit regardless of the
specific loss function employed, and the
$\tau^* \approx 0.85$--$0.95$ threshold reported here may be
interpreted as the sufficiency level for this realistic
class of attenuated forecasts. Rank-perturbation and
Gaussian copula preserve the marginal distribution (exactly
and asymptotically, respectively), and measure a purer
ordinal sufficiency that is less tied to the statistical
signature of pointwise-trained predictors. The slightly
higher $\tau^*$ under these methods reflects this
distinction. A fuller treatment of how $\tau^*$ depends on
the class of forecasts considered lies outside the scope
of this paper.

\section{Results}
\label{sec:results}


\subsection{The Tau-Sufficiency Region}
\label{sec:tau}

\begin{figure}[H]
\centering
\includegraphics[width=0.82\textwidth]{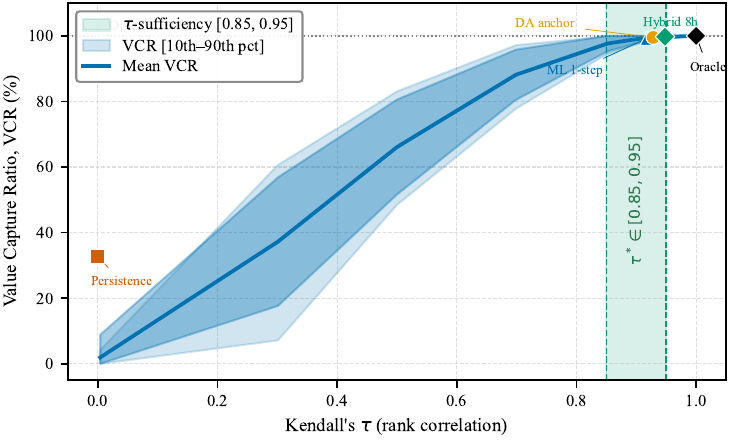}
\caption{Empirical relationship between Kendall $\tau$ and
  Value Capture Ratio (VCR), estimated from 24,000 DP
  simulations with synthetic forecasts of controlled rank
  correlation. The tau-sufficiency region $[0.85, 0.95]$
  (shaded) marks the onset of VCR saturation near 100\%.
  Points indicate the $\tau_K$ and \gls{VCR} values of the
  five benchmark forecasts.}
\label{fig:tau_vcr}
\end{figure}

We begin with the main empirical result of the paper.
Figure~\ref{fig:tau_vcr} shows the relationship between
Kendall $\tauk$ and Value Capture Ratio, estimated from
24,000 \gls{DP} simulations using synthetic forecasts of
controlled rank correlation.

The relationship is highly non-linear.
\gls{VCR} grows steeply as $\tauk$ increases from 0 to 0.70,
with large marginal gains ($\approx$22 percentage points per
0.15 increase in $\tauk$).
Beyond $\tauk \approx 0.85$, the curve enters a plateau:
marginal \gls{VCR} gains fall below 1.4 percentage points
per step, and \gls{VCR} stabilizes near 100\% of oracle
revenue.
We define this plateau as the \emph{tau-sufficiency region}.

\begin{finding}[Tau-Sufficiency Region]
There exists an empirical threshold region
$\tau^* \in [0.85, 0.95]$ in Kendall rank correlation such
that forecasts with $\tauk(f) > \tau^*$ achieve
$\VCR(f) \geq 0.97$, even when absolute forecast errors
remain non-negligible.
Beyond this region, marginal improvements in forecast
accuracy yield negligible economic value.
\end{finding}

Table~\ref{tab:forecast_decision} reports the five benchmark
forecasts evaluated on the 2023--2025 out-of-sample period.
The results confirm the synthetic threshold finding.
Persistence captures only 32.8\% of oracle value, despite
remaining a widely used baseline in electricity forecasting.
By contrast, the ML 1-step forecast captures 99.2\% of oracle
value, and the Hybrid 8h forecast 99.7\%.
The key point is not a complete inversion of the benchmark
ranking, but a strong misalignment between pointwise forecast
error and economic usefulness: persistence may appear
acceptable under simple statistical baselines, yet it
destroys most of the value of the intraday dispatch problem,
whereas forecasts with modest residual error already achieve
near-oracle decisions once they enter the tau-sufficiency
region.

\begin{table}[H]
\centering
\small
\caption{Forecast accuracy vs.\ decision quality for five
  benchmark forecasts. Kendall $\tau$ predicts VCR; MAE does not.
  Persistence achieves the highest \gls{MAE} (30.1~EUR/MWh)
  and the lowest \gls{VCR} (32.8\%), confirming that
  statistical accuracy and decision quality move together
  in extreme cases.
  The more subtle finding is that beyond $\tau^* \approx
  0.85$, further \gls{MAE} reductions yield negligible
  \gls{VCR} improvements: the Hybrid 8h forecast improves
  \gls{MAE} by 53\% over ML 1-step but gains only 0.5
  percentage points in \gls{VCR}.
  Formally, \gls{RI}\,=\,0 across all five benchmarks,
  confirming that the misalignment is quantitative rather
  than ordinal.
  All results are out-of-sample (2023--2025, DE).}
\label{tab:forecast_decision}
\begin{tabularx}{\textwidth}{@{}
    >{\raggedright\arraybackslash}p{2.8cm}
    >{\raggedleft\arraybackslash}X
    >{\raggedleft\arraybackslash}X
    >{\raggedleft\arraybackslash}X
    >{\centering\arraybackslash}p{1.3cm}
    >{\centering\arraybackslash}p{1.3cm}
    @{}}
\toprule
Forecast & MAE & $\tauk$ & VCR (\%) & MAE\,rk & VCR\,rk \\
\midrule
Oracle        & 0.00 & 1.000 & 100.0 & 1 & 1 \\
Hybrid 8h     & 1.30 & 0.948 &  99.7 & 2 & 2 \\
DA + Spread   & 2.22 & 0.928 &  99.5 & 3 & 3 \\
ML 1-step     & 2.75 & 0.918 &  99.2 & 4 & 4 \\
Persistence   & 30.1 & 0.000 &  32.8 & 5 & 5 \\
\bottomrule
\multicolumn{6}{p{0.92\textwidth}}{%
  \footnotesize MAE in EUR/MWh; $\tauk$ = Kendall rank
  correlation with realized prices.
  \textit{Note:} the key misalignment is not a ranking
  inversion across all benchmarks, but the large gap
  between Persistence ($\tau \approx 0$, VCR\,=\,32.8\%)
  and ML-class forecasts ($\tau > 0.9$, VCR\,$>$\,99\%)
  despite similar nominal performance in stable market
  conditions.
  Formally, RI\,=\,0 across all five benchmarks,
  confirming that the misalignment is quantitative
  rather than ordinal.
} \\
\end{tabularx}
\end{table}

The explanation follows directly from the structural argument
of Section~\ref{sec:framework}.
Persistence imports yesterday's intraday shape into today.
When the realized intraday ordering changes, the forecast
provides effectively no useful rank information for the
current day's \gls{DP} problem, regardless of how close its
absolute values are to realized prices.
The ML forecast, by contrast, captures the shape of the
intraday price curve with sufficient fidelity to preserve the
ranking of profitable trading opportunities, even when
individual price levels are imprecise.

Within the tau-sufficiency region, further forecast
improvements yield negligible additional value.
The Hybrid 8h forecast achieves a $\tauk$ of 0.948 and a
\gls{VCR} of 99.7\%---an improvement of only 0.5 percentage
points over the ML 1-step forecast despite substantially
lower \gls{MAE}.
For a battery generating approximately \euro{}1{,}750 per day,
this corresponds to less than \euro{}9 per day of additional
value from a more accurate forecast.
The practical implication is direct: once a forecast achieves
tau-sufficiency, investment in further accuracy improvements
is economically unjustified for intraday dispatch.

The threshold is robust across market conditions.
Repeating the synthetic experiment separately for normal years
(2020--2021, 2024--2025) and crisis years (2022--2023), the
tau-sufficiency region remains $[0.85, 0.95]$ in both
subsamples.
Splitting the out-of-sample weeks by \gls{XBID} volatility,
the $\tauk$ of ML-based forecasts remains stable at
0.929--0.934, comfortably above the threshold in both
high- and low-volatility regimes.
Persistence \gls{VCR}, by contrast, drops by 13.9 percentage
points in high-volatility weeks relative to low-volatility
weeks, confirming that the \gls{MAE} advantage of persistence
is most misleading precisely when arbitrage opportunities
are largest.
Having established the forecast quality threshold for intraday
dispatch, we turn to the capacity allocation decision that
determines how much \gls{XBID} budget is available in the
first place.

\subsection{FCR Dominance, Attribution, and Robustness Frontier}
\label{sec:fcr}

Table~\ref{tab:attribution} reports revenue and \gls{VCR} for
the six configurations of the ablation study over the
2023--2025 out-of-sample period.
The results reveal a clear hierarchy of value sources.
The naive \gls{DA} spread strategy captures only 30.9\% of
oracle value.
Adding Layer~3 \gls{DP} dispatch to a standalone battery with
no reserve obligations achieves $\VCR = 100\%$ relative to
its own \gls{XBID} oracle, confirming that the Rolling
Intrinsic algorithm is near-optimal given its \gls{SoC}
budget.
The dominant value driver is Layer~1 reserve allocation:
introducing \gls{FCR} and \gls{aFRR} participation raises
average daily revenue from \euro{}492 to \euro{}1{,}696,
reflecting the dominant contribution of reserve capacity
payments in the German market during 2023--2025.

\begin{table}[H]
\centering
\caption{Revenue attribution by system configuration
  (2023--2025, DE, out-of-sample).
  Each row adds one decision component.
  \gls{VCR} is computed relative to a configuration-specific
  oracle (perfect foresight under identical market
  participation); no single oracle row is shown because
  each configuration uses a distinct upper bound.
  The oracle revenue for the full multi-market system
  (Layer~1 HMM+SAC) is \euro{}2{,}573/day.
  The HMM and SAC components operate as an integrated
  adaptive bidding system; no separate ablation between
  HMM and SAC is available.}
\label{tab:attribution}
\small
\begin{tabular}{lrrrr}
\toprule
Configuration & Rev./day & VCR (\%) & $\Delta$VCR & CVaR 5\% \\
\midrule
Naive DA spread      &   492 &  46.8 & ---  &   0 \\
DP only (Layer 3)    & 1,432 & 100.0$^\dagger$ & +53.2 & 398 \\
Layer 2\,$+$\,3\,$^{\dag\dag}$ &   565 &  53.7 & ---  &  36 \\
Layer 1 static (Q40) & 1,696 &  64.3 & ---  &  64 \\
Layer~1 HMM+SAC (full system) & 1{,}753 &  68.1 & $+$3.8 &  95 \\
\midrule
Reserve only (static)& 1,131 &  54.5 & ---  &   0 \\
\bottomrule
\multicolumn{5}{l}{\small $^\dagger$ VCR relative to
  oracle\_xbid\_only (standalone XBID battery).}\\
\multicolumn{5}{l}{\small Revenue in EUR/day (mean 2023--2025).
  CVaR 5\% = negative weekly revenue at 5th percentile.}\\
\multicolumn{5}{l}{\footnotesize $^{\dag\dag}$Shown for completeness;
  not included in the Figure~\ref{fig:attribution} ablation chain.}\\
\end{tabular}
\end{table}

The \gls{FCR} allocation is endogenously fixed at 5~\gls{MW}
in all 157 out-of-sample weeks, reflecting the PICASSO buffer
constraint, which limits maximum \gls{FCR} participation to
50\% of power capacity for a 1-hour battery under worst-case
activation assumptions.
The optimizer saturates \gls{FCR} in every week, confirming
that \gls{FCR} capacity is consistently binding in the
modeled system.
The remaining 5~\gls{MW} is allocated dynamically between
\gls{aFRR} (mean 0.65~\gls{MW}, max 3.0~\gls{MW}) and
\gls{XBID} (mean 4.38~\gls{MW}).
Removing the minimum \gls{XBID} participation floor yields
negligible revenue impact (+0.26\% over three years),
confirming that the result is not driven by the participation
constraint.%
\footnote{Without the floor, the optimizer allocates a minimum
of 0.7~\gls{MW} to \gls{XBID} in all weeks, never choosing
zero intraday participation.
The floor is binding in 34 of 157 weeks (21.7\%), with a
revenue impact of \euro{}8k over three years.}

The residual \gls{SoC} budget available to Layer~3 after
fulfilling reserve obligations generates an average of 0.86
equivalent \gls{XBID} cycles per day, following a trimodal
discrete distribution with modes at 0.44, 0.73, and 1.03
cycles per day.
At this cycling intensity, \gls{FCR} marginal revenue is
6.5$\times$ higher per \gls{MW} than \gls{XBID} expected
revenue under identical capacity constraints.
This ratio reflects the physical constraints imposed by
co-participation in reserve markets, not a structural
advantage of \gls{FCR} over intraday arbitrage:
\citet{hornek2025} show that a standalone battery optimized
for continuous intraday trading achieves revenues on the
order of \euro{}150k per \gls{MW}-year in the
German intraday market, reaching levels within the range of \gls{FCR} capacity
revenues reported in the literature; \citet{schaurecker2025} further show that
order-book-level execution yields 14\% higher revenues than
minute-level reoptimization and 58\% higher than hourly
reoptimization in their 2021 German intraday backtest
\citep{schaurecker2025}.
The 6.5$\times$ ratio therefore characterizes multi-market
systems under reserve constraints, with an effective ratio
closer to 2--4$\times$ under optimal standalone execution.

The HMM+SAC adaptive bidding system---in which the
\gls{HMM} identifies the market regime and the SAC
agent selects the optimal bid percentile for that
regime---contributes \euro{}62{,}547 over three years
relative to the static $Q_{40}$ strategy (+3.4\%),
driven by \gls{FCR} acceptance rate improvement from
34\% to 58\% in 2023.
The static $Q_{40}$ strategy, calibrated on the 2022 energy
crisis, submits bids too high for the post-crisis
normalization of 2023.
The HMM identifies the regime shift and adjusts the bid
percentile downward, recovering acceptance rate at a lower
per-\gls{MW} price.
The revenue decomposition confirms the mechanism: \gls{FCR}
revenue increases by +14.8\% under HMM adaptation, while
\gls{aFRR}$+$ revenue declines by 31.5\% as capacity is
redirected toward \gls{FCR}.
The net effect is positive because the \gls{FCR} gain exceeds
the \gls{aFRR} loss in absolute terms.

\begin{figure}[H]
\centering
\includegraphics[width=0.82\textwidth]{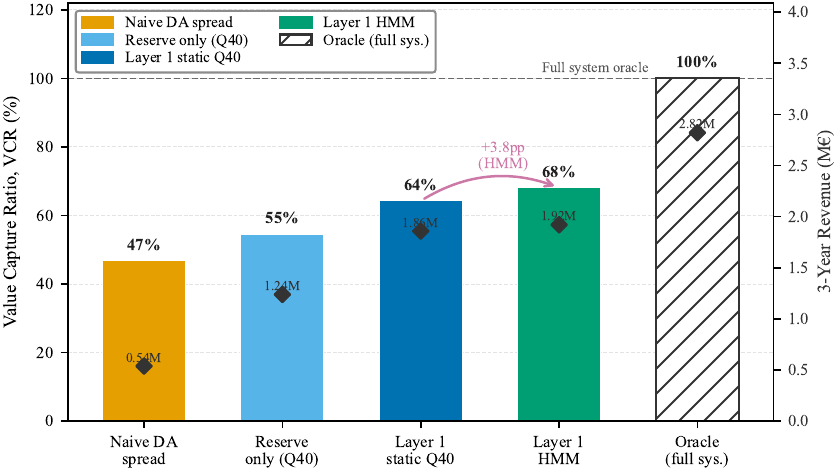}
\caption{Value Capture Ratio (VCR) for each system configuration
  in the ablation study (out-of-sample 2023--2025, DE).
  Each configuration adds one decision layer.
  FCR/aFRR reserve layers contribute the majority of revenue;
  the XBID Layer 3 captures 15--18\% of standalone oracle value
  due to SoC constraints from concurrent reserve obligations.}
\label{fig:attribution}
\end{figure}

Table~\ref{tab:robustness} reports the sensitivity of the
zero-missed-allocation result to higher intraday
monetization potential, tested along two dimensions.
Under Test~A (cycling intensity), the result holds at
0.86 cycles per day (measured mean), 1.03 cycles per day
(measured p90), and 1.30 cycles per day (the enspired
portfolio benchmark).
The frontier is reached at 2.0 cycles per day, where one
week out of 156 would represent a missed allocation---a
48.8\% margin above the empirically observed cycling
intensity.
Under Test~B (price granularity), 15-minute \gls{XBID}
prices yield a spread enrichment of only 1.02--1.05$\times$
over hourly aggregates, confirming that the result is not
an artifact of temporal aggregation.

\begin{table}[H]
\centering
\caption{Robustness frontier for the zero-missed-allocation
  result. The p90--p10 spread is used as the primary robust
  measure. The system operates at 0.86 cycles/day (measured),
  well below the frontier at 2.0 cycles/day.}
\label{tab:robustness}
\begin{tabular}{lrrrr}
\toprule
Scenario & Cycles/day & Spread & FCR/XBID & Missed (\%) \\
\midrule
System (measured mean) & 0.86$^*$ & p90--p10 &
  6.5$\times$ & 0.0 \\
System (measured p90)  & 1.03$^*$ & p90--p10 &
  5.4$\times$ & 0.0 \\
Enspired benchmark     & 1.30     & p90--p10 &
  4.3$\times$ & 0.0 \\
Robustness frontier    & 2.00     & p90--p10 &
  2.8$\times$ & 0.6 \\
\midrule
\multicolumn{5}{l}{\small $^*$ Measured from Layer 3 backtest
  (2023--2025). Trimodal distribution: modes at}\\
\multicolumn{5}{l}{\small 0.44, 0.73, 1.03 cycles/day
  (see Section~\ref{sec:fcr}).}\\
\bottomrule
\end{tabular}
\end{table}

These findings jointly characterize the role of forecast
quality in multi-market \gls{BESS} systems.
Because \gls{FCR} capacity is saturated by the PICASSO
buffer constraint and the residual intraday budget is small
relative to total system capacity, the capacity allocation
decision of Layer~1 is the primary driver of total revenue.
The tau-sufficiency result of Section~\ref{sec:tau} completes
the picture: for the intraday budget that is available,
forecasts above $\tau^* \approx 0.85$ already capture
near-oracle value, making further forecast improvements of
limited incremental value for intraday dispatch decisions.

\subsection{The Swiss Hydro Natural Experiment}
\label{sec:ch}

To test the generality of our findings beyond the German
market, we turn to the Swiss balancing market, which provides
a natural experiment for identifying the role of exogenous
structural factors in \gls{BESS} revenue.
Unlike the German market, where revenue is dominated by
\gls{FCR} capacity payments that are relatively stable across
seasons, the Swiss SRL$-$ market exhibits extreme price
variability driven by hydrological conditions.
When reservoir levels are anomalously high relative to seasonal
norms, the TSO requires additional downward flexibility to
absorb surplus hydraulic generation, driving SRL$-$ prices
sharply higher.

To identify this mechanism empirically, we construct a seasonal
anomaly z-score from weekly Swiss reservoir level data,
adjusting for the strong seasonal pattern in Swiss hydrology.
The running variable is orthogonal to seasonality by
construction, isolating the exogenous variation in reservoir
conditions from regular winter drawdown and summer refill
cycles.

Table~\ref{tab:ch_hydro} reports \gls{BESS} revenue by
hydrological regime.
During high-hydro weeks ($z > 0.7$), weekly SRL$-$ revenue
averages \euro{}20{,}022/\gls{MW}, compared to
\euro{}2{,}480/\gls{MW} during low-hydro weeks ($z < -0.8$)
--- a difference of $+708\%$.
The association is statistically significant:
a one-standard-deviation increase in the anomaly is associated
with approximately \euro{}6{,}400 of additional weekly revenue
per \gls{MW} (OLS, $p = 0.0005$), and the SRL$-$ price
difference between HIGH and LOW regimes is itself highly
significant (Spearman $p = 0.0002$).
The explanatory power is limited ($R^2 = 0.043$), as expected
given the multiple drivers of weekly revenue.

\begin{table}[H]
\centering
\caption{Swiss balancing market revenue by hydrological regime
  (2020--2025). The running variable is the hydrological
  anomaly z-score (seasonally adjusted).
  HIGH\_HYDRO corresponds to $z > 0.7$ standard deviations
  above the seasonal mean.}
\label{tab:ch_hydro}
\begin{tabular}{lrrrr}
\toprule
Regime & N (weeks) & SRL$-$ price & SRL$-$ rev./week & DA price \\
       &           & (EUR/MW/h)  & (EUR/MW)         & (EUR/MWh) \\
\midrule
LOW\_HYDRO ($z < -0.8$)  &  78 & 16.2 &  2,480 & 108.2 \\
MEDIUM                    & 157 & 16.7 &  4,374 & 110.2 \\
HIGH\_HYDRO ($z > 0.7$)   &  78 & 63.7 & 20,022 & 149.1 \\
\midrule
\multicolumn{5}{l}{HIGH vs LOW: $\Delta$SRL$-$ rev.\ $=+708\%$;
  price difference $p=0.0002$ (Spearman).}\\
\multicolumn{5}{l}{OLS total revenue: coeff $=6{,}369$ EUR/SD,
  $R^2=0.043$, $p=0.0005$.}\\
\bottomrule
\end{tabular}
\end{table}

The low $R^2$ is expected.
Weekly \gls{BESS} revenue depends on multiple drivers ---
\gls{DA} prices, gas prices, \gls{aFRR} dynamics --- of which
hydrological anomaly is only one.
High statistical significance combined with low explanatory
power is therefore consistent with a real but partial effect,
as is standard in empirical market studies using
macroeconomic running variables.

The lead-lag structure of the relationship provides additional
evidence supporting the proposed mechanism.
Spearman correlation between the hydro anomaly and SRL$-$
prices is strongest at a lag of two to four weeks, suggesting
that reservoir conditions are observable in advance and that
price effects materialize with a delay as the TSO adjusts
procurement volumes.
This lead-lag structure is consistent with exogenous
hydrological variation driving the price signal rather than
the reverse.

\begin{figure}[H]
\centering
\includegraphics[width=\textwidth]{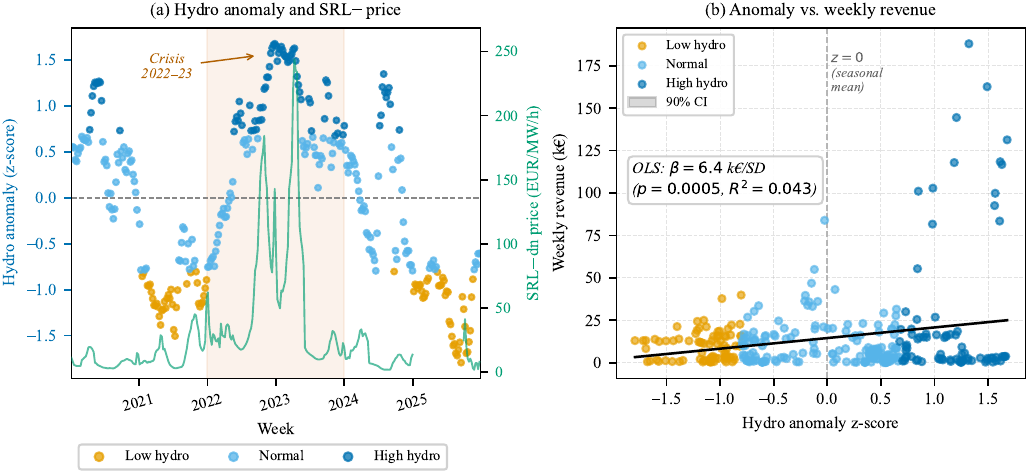}
\caption{Relationship between Swiss hydrological anomaly
  and SRL$-$ balancing market revenue (2020--2025).
  Left: time series of hydrological anomaly z-score (blue)
  and SRL$-$dn capacity prices (green). Right: scatter plot of seasonal
  anomaly z-score vs.\ weekly BESS revenue from SRL$-$,
  with OLS fit ($R^2 = 0.043$, $p = 0.0005$).
  The 2022--2023 storage scarcity crisis (Speicherkrise)
  is annotated.}
\label{fig:ch_hydro}
\end{figure}

We interpret these results as evidence consistent with an
exogenous hydrological shock, without making a strict causal
claim.
The identification strategy relies on the orthogonality of the
seasonal anomaly to regular market dynamics; potential
confounders include correlated European weather events that
simultaneously affect both reservoir levels and wholesale
electricity prices.
The finding nonetheless provides the first empirical
documentation of the hydro-balancing market transmission
mechanism in Switzerland, with direct implications for
\gls{BESS} operators in hydro-dependent markets such as
Austria, Norway, and Iberia, where similar structural drivers
may shape balancing market revenues.

\section{Discussion}
\label{sec:discussion}

The three empirical findings of Section~\ref{sec:results}---
the tau-sufficiency threshold, the \gls{FCR} dominance under
reserve constraints, and the Swiss hydro transmission
mechanism---jointly reframe how \gls{BESS} operators should
evaluate forecast quality and allocate system capacity.
We discuss their theoretical and practical implications in turn.

\subsection{Theoretical Implication}
\label{sec:theory}

The tau-sufficiency result has a theoretical implication that
extends beyond battery storage applications.
The \gls{DP} Rolling Intrinsic algorithm is a specific instance
of a broader class of optimization problems in which the
objective function depends on the relative ordering of input
parameters rather than their absolute values.
A broad class of such problems---including certain forms of
inventory management, scheduling, and trading---share this
ordinal structure: the optimal policy is determined by the
ranking of future states, not their precise magnitude.
For any such problem, standard pointwise error metrics are
misaligned with decision quality by construction, and rank
correlation metrics are more appropriate evaluation criteria.

This connects to the emerging literature on decision-focused
learning \citep{smets2024, vanderschueren2022}, which proposes
retraining forecast models to minimize downstream decision cost
rather than statistical error.
While decision-focused learning has been shown to improve
out-of-sample profits for energy storage systems, albeit
at the cost of significant computational complexity
\citep{smets2024}, this approach integrates the downstream
objective into model training rather than characterizing
the decision structure itself.
Our finding is complementary but distinct: we show that for
problems with an ordinal decision structure, the alignment
between statistical metrics and decision quality can be empirically
characterized by a single threshold in rank correlation.
This threshold is a property of the decision problem, not of
the forecasting method, and is therefore stable across forecast
architectures.
For tau-sufficient forecasts, decision-focused retraining may
offer limited additional value---a result that simplifies the
operational pipeline for \gls{BESS} operators.
Consistent with \citet{vanderschueren2022}, who show
empirically that decision-making strategies have a larger
impact on cost performance than the objective function
used for training, our results highlight that improvements
in the decision framework can outweigh gains from
modifying the predictive model itself.

\subsection{Practical Implications for BESS Operators}
\label{sec:practical}

The findings translate into three concrete operational
recommendations.

First, forecast evaluation for intraday dispatch should
prioritize rank-based metrics over pointwise error metrics
such as \gls{MAE} or \gls{RMSE}.
The relevant question for a \gls{BESS} operator is not
``what is my forecast error?'' but ``does my forecast preserve
the ranking of intraday price opportunities?''
In our setting, forecasts achieving $\tauk > 0.85$ are
tau-sufficient: further investment in forecast accuracy will
not materially improve dispatch revenue.

Second, investment in forecast improvement should prioritize
Layer~1 over Layer~3.
Our attribution results show that the capacity allocation
decision dominates total revenue, while intraday dispatch
quality is a second-order effect once tau-sufficiency is
achieved.
Improving the cross-market spread forecast that informs
Layer~1 allocation---particularly during regime transitions
when the static $Q_{40}$ strategy underperforms---has higher
expected return than improving the intraday price forecast.

Third, operators in hydro-dependent markets should incorporate
reservoir level monitoring into their bid strategy.
The Swiss evidence shows that hydrological anomalies are
observable two to four weeks in advance and are associated
with SRL$-$ price spikes that can increase weekly revenue by
more than 700\%.
A simple regime-switching strategy that raises bid percentiles
during high-hydro periods---analogous to the HMM mechanism
applied to \gls{FCR} in the German market---could capture a
significant fraction of this opportunity.

\subsection{Market Regime and Heterogeneity}
\label{sec:regime}

The findings should be interpreted in the context of the
German and Swiss markets over 2020--2025.
Two sources of heterogeneity deserve attention.

The \gls{FCR} dominance result depends on the current
structure of the German balancing market, where \gls{BESS}
penetration in \gls{FCR} has grown rapidly but has not yet
saturated to the point of driving capacity prices to zero.
As \gls{BESS} capacity continues to expand, \gls{FCR} prices
may decline as \gls{BESS} penetration increases, compressing
the \gls{FCR}/\gls{XBID} revenue ratio.
The tau-sufficiency result, by contrast, is structural: it
depends on the ordinal nature of the \gls{DP} dispatch
problem, not on the relative profitability of \gls{FCR} and
\gls{XBID}, and will remain valid as market conditions evolve.

The Swiss hydro finding highlights the importance of
market-specific structural drivers that are absent from
German-focused models.
Similar mechanisms may operate in other hydro-dependent
European markets---Austria, Norway, and Iberia---where
hydroelectric generation plays a dominant role in reserve
provision.
Operators and researchers working in these markets should
account for hydrological conditions as a primary revenue
driver for downward balancing services, rather than treating
balancing prices as purely driven by gas and demand dynamics.

\subsection{Limitations and Scope}
\label{sec:limitations}

Three limitations of this study should be acknowledged.

First, \gls{aFRR} energy activations are modeled via
expected value and \gls{SoC} buffer constraints rather
than fully dynamic real-time simulation.
This simplification is expected to be conservative for
revenue estimation: the PICASSO buffer is calibrated to
worst-case activation scenarios, so any dynamic
activation exceeding the buffer would reduce available
\gls{XBID} capacity and bias revenue estimates downward
rather than upward.
However, the dynamic coupling between \gls{aFRR}
activations and \gls{XBID} decisions is not fully
modeled.
In practice, concurrent activations reduce \gls{XBID}
capacity in real time, creating interactions that the
layer-separated architecture does not capture.
This may result in occasional infeasibility in
high-activation periods that the static buffer approach
does not detect.

Second, \gls{XBID} trading is modeled as price-taking at
15-minute aggregated prices from SMARD, abstracting from
order book dynamics, bid-ask spreads, and partial fills.
\citet{schaurecker2025} demonstrate that order-book-level
execution yields substantially higher intraday revenues than
aggregated-price models, implying that our \gls{XBID} revenue
estimates are conservative.
The \gls{FCR} dominance ratio of 6.5$\times$ would be closer
to 2--4$\times$ under optimal intraday execution, as discussed
in Section~\ref{sec:fcr}.

Third, the tau-sufficiency threshold
$\tau^* \approx 0.85$--$0.95$ is estimated on German
\gls{XBID} data for 2023--2025.
While the existence of a threshold is a structural
property of ordinal decision problems, its precise
numerical value may vary across markets with different
price dynamics, liquidity structures, or trading
horizons.
Validation on other European markets would refine the
quantitative estimate of $\tau^*$ and strengthen the
generalizability of the finding.

\section{Conclusion}
\label{sec:conclusion}

This paper examines the relationship between forecast accuracy
and decision quality in multi-market battery storage
optimization, using six years of real market data from Germany
and Switzerland.

Three findings emerge.
First, we identify an empirical tau-sufficiency region
($\tau^* \approx 0.85$--$0.95$) in Kendall rank correlation
beyond which value capture saturates near 100\% of oracle
revenue, with limited sensitivity to further improvements in
absolute forecast error.
The \gls{DP} dispatch problem is inherently ordinal---it
depends on the relative ranking of prices, not their absolute
levels---and any forecast that sufficiently preserves this
ranking induces a near-optimal policy.
Standard accuracy metrics such as \gls{MAE} fail to capture
this property, leading to systematic misalignment between
statistical evaluation and economic performance.

Second, in multi-market systems under reserve constraints,
the capacity allocation decision of Layer~1 is the primary
driver of total revenue.
\gls{FCR} capacity is endogenously saturated by the PICASSO
buffer constraint in all observed weeks of our sample, making
the Layer~1 allocation---not the intraday forecast---the
economically dominant decision.
The tau-sufficiency result completes the picture: for the
intraday budget that remains after reserve obligations,
standard \gls{ML} forecasts already achieve tau-sufficiency,
making further forecast improvement of limited incremental
value for intraday dispatch.

Third, the Swiss balancing market provides one of the first
empirical documentations of the hydro-balancing market
transmission mechanism: hydrological surplus anomalies are
significantly associated with balancing market prices
(Spearman $p = 0.0002$) and revenue (OLS $p = 0.0005$),
with high-hydro weeks generating
approximately 8$\times$ higher downward reserve revenue than
low-hydro weeks.
This finding highlights the importance of market-specific
structural drivers that are absent from German-focused models
and relevant for \gls{BESS} operators in hydro-dependent
markets across Europe.

Together, these results reframe the forecast evaluation
problem for \gls{BESS} operators.
The relevant question is not ``what is my \gls{MAE}?'' but
``does my forecast achieve tau-sufficiency, and is my
capacity allocation strategy adapted to the current market
regime?''
Rank correlation metrics and regime-aware bidding, rather
than pointwise accuracy and static strategies, are the
primary tools that deliver economic value in heterogeneous
European balancing markets.

\appendix

\section{HMM+SAC Adaptive Bidding: Technical Details}
\label{app:hmm}

We implement a Hidden Markov Model following
the standard formulation of \citet{rabiner1989},
using four observable features computed
weekly: the rolling mean \gls{FCR} clearing price (4-week
window), its standard deviation, the week-over-week price
change, and the \gls{aFRR} acceptance rate.
These features are standardized before training.
The \gls{HMM} is trained via Baum-Welch on the full
2020--2022 in-sample period and applied in a walk-forward
fashion from 2023 onward, with the state sequence
re-estimated monthly.

The four latent states correspond to identifiable market
regimes: normal (\gls{FCR} $\approx$\,16 EUR/\gls{MW}/h),
crisis (\gls{FCR} $\geq$\,20 EUR/\gls{MW}/h), post-crisis
normalization (\gls{FCR} declining rapidly), and
low-volatility (\gls{FCR} stable at low levels).
Regime assignment is obtained from the most
likely latent state sequence under the fitted
model, using the Viterbi algorithm.

The Soft Actor-Critic (SAC) agent \citep{haarnoja2018} receives the current
\gls{HMM} state as input and outputs a bid percentile
$p \in [Q_{20}, Q_{60}]$, trained to maximize realized
weekly \gls{FCR} revenue.
Training uses 2020--2022 data with a replay buffer of
52 weeks and a discount factor $\gamma = 0.95$.
The policy is fixed from 2023 onward (no online learning
in the out-of-sample period).
Validation on a held-out 2022\,H2 window confirmed that
the policy improves \gls{FCR} acceptance rate relative to
the static $Q_{40}$ strategy in all four regime states,
with the largest gain in the post-crisis normalization
state (acceptance rate: 34\%\,$\to$\,58\%).
Out-of-sample performance is stable across all four
regimes, with no evidence of overfitting in the
held-out 2022\,H2 validation window.

The \gls{HMM}+SAC system adds \euro{}62{,}547 over three
out-of-sample years relative to the static $Q_{40}$
baseline (+3.4\%), with the improvement concentrated in
the 2023 regime transition period.
The module is modular and can be replaced with any
regime-detection mechanism without affecting the Layer~1
LP structure or the Layer~3 \gls{DP} dispatch.

\section*{Acknowledgements}

This research did not receive any specific grant from
funding agencies in the public, commercial, or
not-for-profit sectors.

\section*{Declaration of Generative AI and AI-Assisted Technologies in the Manuscript Preparation Process}

During the preparation of this work the author used Claude (Anthropic) to assist with manuscript drafting, literature positioning, and \LaTeX{} implementation. The author reviewed and edited all content, verified all numerical results against the underlying codebase and data, and takes full responsibility for the content of the published article.

\bibliographystyle{elsarticle-harv}
\bibliography{references}

\end{document}